\documentclass[prl,reprint,amsmath,amssymb,showpacs,twocolumn]{revtex4}
\usepackage{dcolumn}
\usepackage{color}
\usepackage{amsmath}
\usepackage{epsf}
\usepackage{graphicx}
\usepackage{bm}

\begin{document} 

\preprint{APS/123-QED}
\title{Nonlocal spin correlations mediated by a superconductor}

\author{Taewan Noh$^1$}
\author{Manuel Houzet$^2$, Julia S. Meyer$^2$}
\author{Venkat Chandrasekhar$^1$}
\affiliation{$^1$Department of Physics and Astronomy, Northwestern University, Evanston, IL 60208, USA, \\ 
$^2$ SPSMS, UMR-E 9001 CEA/UJF-Grenoble 1, INAC, Grenoble, F-38054, France}
\date{\today}

\begin{abstract}
Nonlocal charge correlations induced in two normal metals contacted separately to a superconductor have been studied intensively in the past few years. Here we investigate nonlocal correlations induced by the transfer of pure spin currents through a superconductor on a scale comparable to the superconducting coherence length. As with charge currents, two processes contribute to the nonlocal spin signal:  crossed Andreev reflection (CAR), where an electron with spin up injected from one normal metal into the superconductor results in a hole with spin down being injected into the second normal metal, and elastic cotunneling (EC), where the electron with spin up injected from the first normal metal results in an electron with spin up being injected into the second normal metal. Unlike charge currents, however, the spin currents associated with CAR and EC add due to the fact that the bulk superconductor cannot sustain a net spin current.
\end{abstract}


\maketitle

Electrons in two spatially separated normal metals in contact with a superconductor show nonlocal correlations that are mediated by their mutual interaction with the superconductor \cite{Byers, Deutscher, Falci, Bignon}.  
Two processes are responsible for these correlations. In the first process, called crossed Andreev reflection (CAR), an electron with one spin orientation, e.g., spin up, is injected from the first normal metal (N) into the superconductor (S), resulting in a hole with spin down being injected from the superconductor into the second normal metal, with a concomitant generation of a Cooper pair in the superconductor.  In the second process, elastic cotunneling (EC), the spin-up electron injected from the first normal metal gives rise to a spin-up electron being injected into the second normal metal.  Both processes are exponentially suppressed with the distance between the two NS interfaces on a length scale of the order of the superconducting coherence length $\xi_S$ \cite{Feinberg}. 
In the case of \textit{charge} currents, the net current injected into the second normal metal in response to the drive current injected from the first normal metal is the \textit{difference} of the two contributions CAR and EC.  As the relative amplitude of CAR and EC is predicted to depend on the transparency of the interface \cite{Melin}, the effect of electron-electron interactions \cite{Levy-Yeyati},
and so forth, the sign of the net current injected into the second normal metal may be positive or negative.  Experimentally, nonlocal correlations due to CAR and EC in charge transport have been verified by many different groups \cite{Beckmann, Russo, cz}.

In this paper, we study theoretically nonlocal correlations induced in a NSN structure in response to a pure spin current. 
As with charge currents, both CAR and EC contribute to the resulting nonlocal spin signal.  However,  due to the fact that one cannot inject a pure spin current into the bulk of a $s$-wave superconductor (which is the case of interest here), we find that the resulting nonlocal spin current 
is the \textit{sum} of CAR and EC contributions: Injecting a spin-down hole (CAR) or injecting a spin-up electron (EC) into the second normal metal, both correspond to injecting a net up spin. Consequently, with a combination of charge and spin transport measurements on the same device, one should in principle be able to separate the contributions of CAR and EC, which has not been possible to do so far.

In addition to CAR and EC, there are additional processes that may contribute to the nonlocal signals in NSN or FSF structures. These are charge \cite{Clarke, Tinkham, Brauer, Hubler, Arutyunov} and spin \cite{Hubler2, Quay} imbalance, associated with the injection of quasiparticles, with energies larger than the superconducting gap ∆, into the superconductor. In contrast to these studies, the proposed experiment 
addresses the regime of subgap transport. Thus, no quasiparticle is injected above the gap, and the effect does not depend on the long spin relaxation times recently observed in the nonlocal measurements of Refs. \cite{Hubler2, Quay} at applied bias voltage larger than $\Delta$/e.

\begin{figure}
      \begin{center}
      \includegraphics[width=8cm]{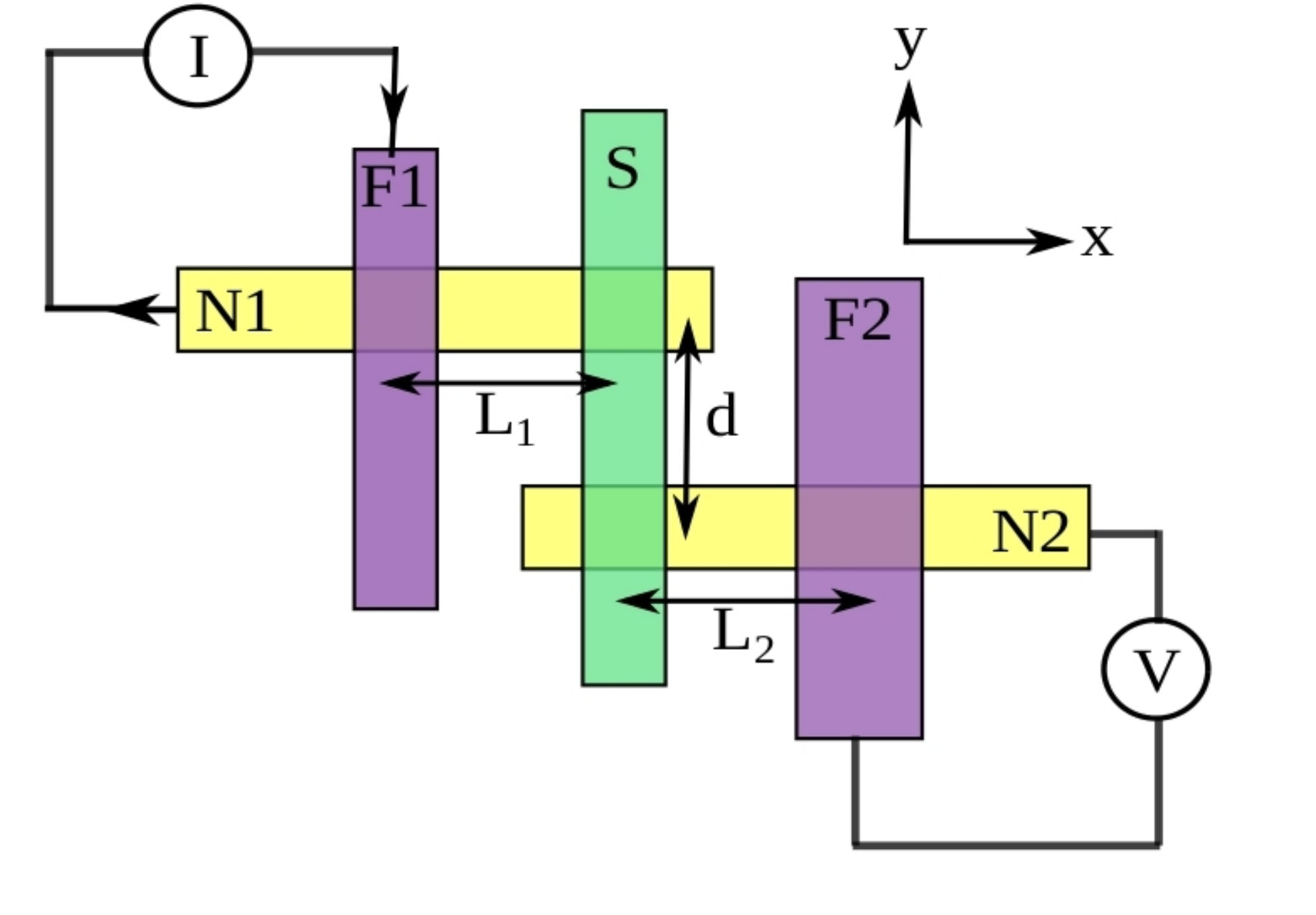}
      \caption{Schematic diagram of the device. The bias current is applied from a ferromagnet F$_1$ to a normal metal
      N$_1$ and nonlocal voltage is measured between F$_2$ and N$_2$.} 
      \label{fig1}
\end{center}
\end{figure}

Figure \ref{fig1} shows the schematic view of the device geometry that we consider here. Our device is similar to the usual NSN hybrid structure, except that there are two additional ferromagnets, F$_1$ and F$_2$, placed in contact with the two normal metals, N$_1$ and N$_2$, respectively.  One of the normal-metal/ferromagnet junctions (N$_1$F$_1$) serves as a spin injector.  If one drives a bias charge current $I$ from F$_1$ to N$_1$, as shown in Fig.~\ref{fig1}, spin accumulates at the N$_1$F$_1$ interface. The resulting spin imbalance leads to spin diffusion in all possible directions in N$_1$.  Since the charge current is drained from the left side of N$_1$, the right side of N$_1$ carries a \textit{pure} spin current with no net flow of charge. The second NF interface (N$_2$F$_2$) is used to detect the spin current flowing through N$_2$ by measuring  the nonlocal voltage $V_\mathrm{nl}$ that establishes itself between N$_2$ and F$_2$. If F$_1$ and F$_2$ are designed to have different coercive fields, it 
is 
possible to realize both parallel and antiparallel magnetization directions by applying an external magnetic field. Below we determine the nonlocal spin signal by computing the nonlocal resistance $R_\mathrm{nl} =V_\mathrm{nl} /I$. 

In the following, we assume that  the normal metals are oriented along the $x$-axis whereas the ferromagnets are oriented along the $y$-axis (see Figure \ref{fig1}). 
To simplify the notation, we use two different coordinate systems for N$_1$ and N$_2$ with origins at the  respective NF interface and the $x$-axis directed toward the superconductor, i.e., the respective NS interfaces are at $x=L_i$ ($i=1,2$).

 \begin{figure}
      \begin{center}
      \includegraphics[width=7.5cm]{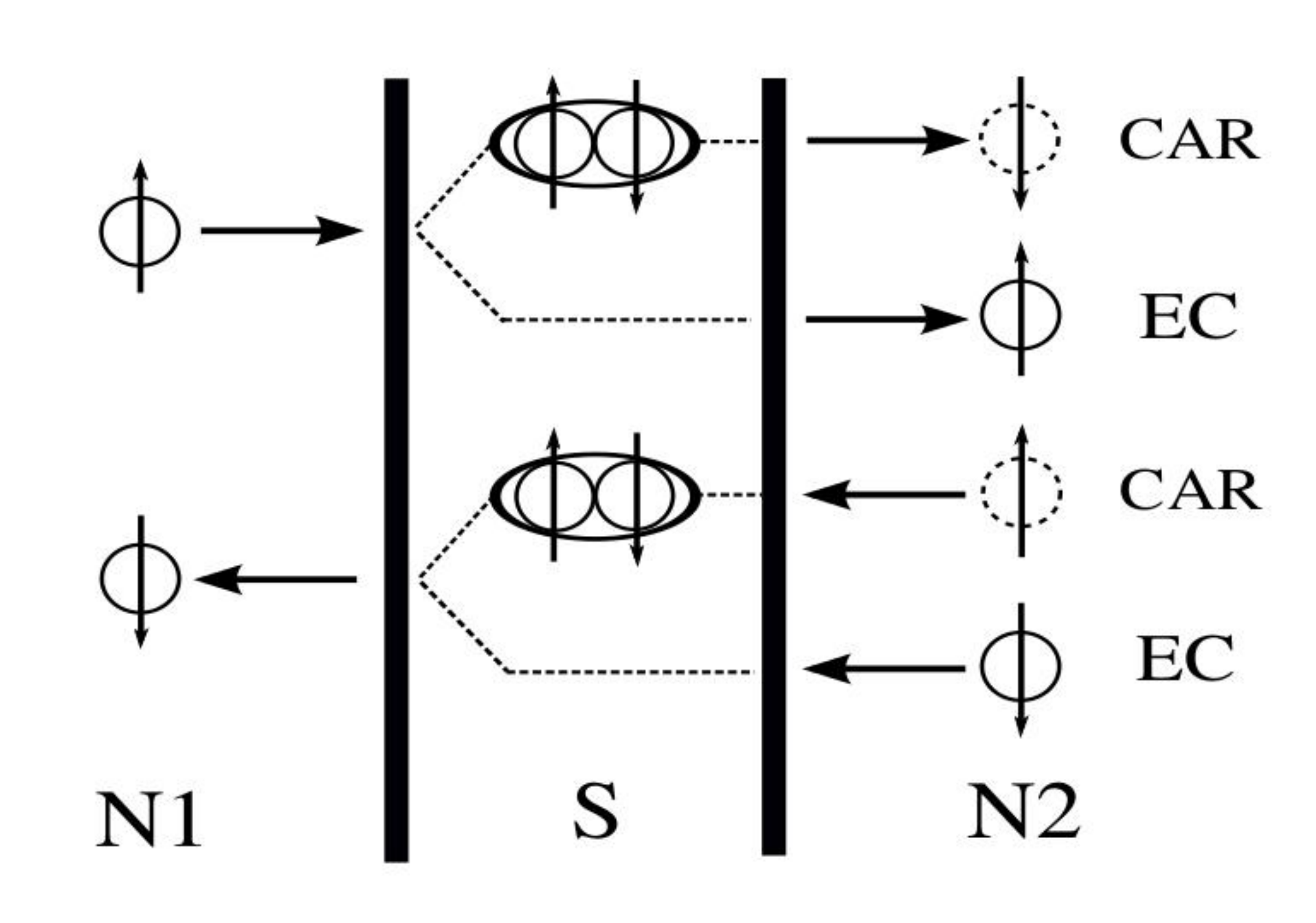}
      \caption{Nonlocal correlations near N$_1$SN$_2$ structure. Pure spin current is injected from N$_1$ and a finite spin current is 
               generated in N$_2$ by CAR and EC. One should note that the direction of the spin currents attributed to CAR and EC
               are in the same direction and there is no net spin accumulation in the superconductor.} 
      \label{fig2}
\end{center}
\end{figure}

Let us first consider the processes that occur at the N$_1$SN$_2$ interfaces in more detail. As shown in Fig.~\ref{fig2}, an incoming electron with subgap energy 
may undergo four possible processes at the N$_1$S interface: local Andreev reflection (AR), normal reflection (NR), crossed Andreev reflection (CAR), and elastic cotunneling (EC). Only the nonlocal processes (CAR and EC) contribute to the nonlocal signal. If the length of the superconducting link between N$_1$ and N$_2$ is much longer than $\xi_S$, these processes are suppressed. In that limit, only AR and NR occur at the N$_1$S interface, and thus the detector voltage between N$_2$ and F$_2$ is zero. 
In the opposite regime, when $d$ is shorter than $\xi_S$, however, nonlocal processes such as CAR and EC can occur and generate a spin current in
N$_2$.  The resulting spin accumulation at the N$_2$F$_2$ interface yields a finite voltage difference between N$_2$ and F$_2$.  It is important to note here that, in the subgap regime, the superconductor cannot support spin accumulation: the spin of an electron injected into the superconductor at the N$_1$S interface is transferred to N$_2$, either via a hole with the opposite spin (CAR) or via an electron with the same spin (EC), thus leaving no net spin in the superconductor. The spin current injected from N$_1$ into S, $I^s_{\mathrm{N}_1\mathrm{S}}$, equals the spin current injected from S into N$_2$,  $-I^s_{\mathrm{N}_2\mathrm{S}}$. Here
$I^s_{\mathrm{N}_i\mathrm{S}}=I^\uparrow_{\mathrm{N}_i\mathrm{S}}-I^\downarrow_{\mathrm{N}_i\mathrm{S}}$ is the difference of spin-up and spin-down currents,
with a fixed spin quantization axis collinear with the magnetizations in F$_{1/2}$.
As a consequence, as can be seen schematically in Fig. \ref{fig2}, both CAR and EC contribute to a spin current in N$_2$ in the same direction, such that their contributions to the spin signal add.  This is completely different from the case of charge current, where one measures the difference between CAR and EC. 
Overall, the spin current through the superconductor, $I^s_{\rm S}=I^s_{\mathrm{N}_1\mathrm{S}}=-I^s_{\mathrm{N}_2\mathrm{S}}$, is related to the difference between the spin imbalances $\delta \mu_{\mathrm{N}_i}(L_i)=[\mu^\uparrow_{\mathrm{N}_i}(L_i)-\mu^\downarrow_{\mathrm{N}_i}(L_i)]/2$ at the two interfaces,
\begin{align}
\label{eq:IsS}
I^s_{\rm S}
=\frac{G_{\rm S}^+}{e}[\delta \mu_{\mathrm{N}_1}(L_1) -\delta \mu_{\mathrm{N}_2}(L_2)] ,
\end{align}
Here $\mu^{\uparrow/\downarrow}_{\mathrm{N}_i}$ are the spin-resolved electrochemical potentials. Note that  
$\mu_{\mathrm{N}_1}(L_1)=\mu_{\mathrm{N}_2}(L_2)$,
with 
$ \mu_{\mathrm{N}_i}(L_i)=[\mu^\uparrow_{\mathrm{N}_i}(L_i)+\mu^\downarrow_{\mathrm{N}_i}(L_i)]/2$, in the absence of a charge current. Furthermore, $G_{\rm S}^+ = G_{\mathrm{CAR}}+G_{\mathrm{EC}}$, where $G_{\mathrm{CAR/EC}}$ are the conductances due to CAR and  EC, respectively.

To describe spin accumulation and spin transport at the NF interfaces, we use the model introduced by Takahashi \textit{et al.} \cite{Taka}. 
The current density ${j}^{\sigma}_\alpha=-(\sigma_\alpha^\sigma/e){\nabla}\mu_\alpha^\sigma$ for spin $\sigma=\uparrow,\downarrow$ is 
related with the spin resolved conductivity $\sigma_\alpha^\sigma$ and electrochemical potential $\mu_\alpha^\sigma$ in material $\alpha=$ F$_i$,N$_i$.
The continuity equation for the charge current density ${j}_\alpha={j}_\alpha^\uparrow+{j}_\alpha^\downarrow$ reads 
\begin{equation}
\label{eq:j}
{\nabla}{j}_\alpha=0,
\end{equation}
while the relaxation of spin imbalance is described by the phenomenological equation 
\begin{equation}
\label{eq:deltamu}
{\nabla}^2\delta \mu_\alpha-(1/\lambda_\alpha^2)\delta \mu_\alpha=0,
\end{equation}
where $\lambda_\alpha$ is the spin diffusion length. The spin current density is ${j}^s_\alpha={j}_\alpha^\uparrow-{j}_\alpha^\downarrow$.

The general solution of Eqs.~\eqref{eq:j} and \eqref{eq:deltamu} in F$_i$ reads
\begin{eqnarray}
\label{eq:muF}
\mu^{\uparrow/\downarrow}_{\mathrm{F}_i}(y)
&=&
\mu_{\mathrm{F}_i}(0)+
\left(p_{\mathrm{F}_i}\pm\frac{2\sigma_{\mathrm{F}_i}^{\downarrow/\uparrow}}{\sigma_{\mathrm{F}}}e^{-|y|/\lambda_\mathrm{F}}\right)\delta\mu_{\mathrm{F}_i}(0)\nonumber\\
&&+\frac{eI_iy}{\sigma_{\mathrm{F}}A_{\mathrm{F}}}\Theta({y}) 
\end{eqnarray}
with $I_1=I$ and $I_2=0$. $\Theta(y)$ is the Heaviside function.  Note that the chemical potentials in the ferromagnets are defined in the same way as in the normal metals, i.e., $\mu_{F_i}^{\uparrow/\downarrow}(y) = \mu_{F_i}(y) \pm \delta \mu_{F_i}(y)$. For simplicity, we assume identical cross-sections $A_\mathrm{F}$, spin-diffusion lengths $\lambda_\mathrm{F}$, and total conductivities $\sigma_{\mathrm{F}}=\sigma^\uparrow_{\mathrm{F}_i}+\sigma^\downarrow_{\mathrm{F}_i}$ in the wires F$_1$ and F$_2$, while $p_{\mathrm{F}_i}=(\sigma^\uparrow_{\mathrm{F}_i}-\sigma^\downarrow_{\mathrm{F}_i})/\sigma_{\mathrm{F}}$.
In particular, $p_{\mathrm{F}_1}$ and $p_{\mathrm{F}_2}$ have the same modulus $p_{\mathrm{F}}$, while their sign is determined by the magnetization direction in F$_i$, $p_{\mathrm{F}_i}=\pm p_{\mathrm{F}}$, depending on whether  
electrons with spin up/down correspond to majority/minority ($+$) or minority/majority ($-$) electrons, respectively.
Using Eq. \eqref{eq:muF} to compute the current densities at $y=0^\pm$, we can relate the spin imbalance at the FN interfaces to the drive current and the spin currents $I^s_{\mathrm{F}_i\mathrm{N}_i}$ injected from F$_i$ to N$_i$ through the F$_i$N$_i$ interface,
\begin{equation}
\label{eq:muF0}
\delta \mu_{\mathrm{F}_i}(0)=
\frac12eR_\mathrm{F}\left(p_{\mathrm{F}_i} I_i-I^s_{\mathrm{F}_i\mathrm{N}_i}\right),
\end{equation}
where $R_\mathrm{F}=\lambda_{\mathrm{F}}/[A_{\mathrm{F}}\sigma_{\mathrm{F}}(1-p^2_{\mathrm{F}})]$.

The currents through the  F$_i$N$_i$ interface may also be expressed in terms of the potential drops across the interface,
\begin{subequations}
\label{eq:GT}
\begin{eqnarray}
I_i
&=&
\frac{G_{T_i}}e
\left(\mu_{\mathrm{F}_i\mathrm{N}_i}+p_{T_i}\delta\mu_{\mathrm{F}_i\mathrm{N}_i}\right),
\\
I^s_{\mathrm{F}_i\mathrm{N}_i} 
&=&
\frac{G_{T_i}}e
\left(p_{T_i}\mu_{\mathrm{F}_i\mathrm{N}_i}+\delta\mu_{\mathrm{F}_i\mathrm{N}_i}\right),
\end{eqnarray}
\end{subequations}
where $G_{T_i}= G^{\uparrow}_{T_i} +G^{\downarrow}_{T_i}$ and 
$p_{T_i} =(G^{\uparrow}_{T_i} -G^{\downarrow}_{T_i})/G_{T_i}$
are related with the (tunnel) conductances $G^{\uparrow/\downarrow}_{T_i}$ of the N$_i$F$_i$ interface for spin up and spin down electrons. 
Furthermore, $\mu_{\mathrm{F}_i\mathrm{N}_i}=\mu_{\mathrm{F}_i}(0)-\mu_{\mathrm{N}_i}(0)$ and $\delta \mu_{\mathrm{F}_i\mathrm{N}_i}=\delta \mu_{\mathrm{F}_i}(0)-\delta \mu_{\mathrm{N}_i}(0)$.
The sign of $p_{T_i}$ again depends on the magnetization direction of the ferromagnet. Note that spin-flip scattering at the interface has been neglected in Eqs.~\eqref{eq:GT} for simplicity. Inverting these equations, we obtain 
\begin{subequations}
\label{eq:GT2}
\begin{eqnarray}
\mu_{\mathrm{F}_i\mathrm{N}_i}
&=&
eR_{T_i}
\left(I_i-p_{T_i}I^s_{\mathrm{F}_i\mathrm{N}_i} \right),
\label{eq:GT2a}
\\
\delta\mu_{\mathrm{F}_i\mathrm{N}_i} 
&=&
eR_{T_i}
\left(-p_{T_i}I_i+I^s_{\mathrm{F}_i\mathrm{N}_i} \right),
\label{eq:GT2b}
\end{eqnarray}
\end{subequations}
where $R_{T_i}=1/[G_{T_i}(1-p_{T_i}^2)]$.

The general solution of Eqs.~\eqref{eq:j} and \eqref{eq:deltamu} in the normal metals N$_i$ reads
\begin{subequations}
\label{eq:muNx}
\begin{eqnarray}
\mu_{\mathrm{N}_i}(x)
&=&
\mu_{\mathrm{N}_i}(0)-\frac{eI_ix}{\sigma_{\mathrm{N}}A_{\mathrm{N}}}\Theta(-x),
\\
\delta\mu_{\mathrm{N}_i}(x)
&=&
\frac 1 2 eR_\mathrm{N}
\left(I^s_{\mathrm{F}_i\mathrm{N}_i} 
e^{-\frac{|x|}{\lambda_{\mathrm{N}}}}
-I^s_{\mathrm{N}_i\mathrm{S}_i} e^{-\frac{|L_i-x|}{\lambda_{\mathrm{N}}}}\right),
\end{eqnarray}
\end{subequations}
with $R_\mathrm{N}\equiv 1/G_\mathrm{N}={\lambda_{\mathrm{N}}}/{\sigma_{\mathrm{N}}A_{\mathrm{N}}}$. As for the ferromagnets, here we also assume identical cross-sections $A_\mathrm{N}$ and spin-diffusion lengths $\lambda_\mathrm{N}$, and conductivities $\sigma_N$ in the wires N$_1$ and N$_2$. Furthermore, $\sigma^\uparrow_{\mathrm{N}}=\sigma^\downarrow_\mathrm{N}=\sigma_{\mathrm{N}}/2$. Using these equations, we find in particular
\begin{subequations}
\label{eq:muN}
\begin{eqnarray}
\delta\mu_{\mathrm{N}_i}(0)
&=&
\frac12 eR_\mathrm{N}
\left(I^s_{\mathrm{F}_i\mathrm{N}_i} 
-I^s_{\mathrm{N}_i\mathrm{S}_i} e^{-L_i/\lambda_{\mathrm{N}}}\right),
\\
\delta\mu_{\mathrm{N}_i}(L_i)
&=&
\frac12 eR_\mathrm{N}
\left(I^s_{\mathrm{F}_i\mathrm{N}_i} e^{-L_i/\lambda_{\mathrm{N}}}
-I^s_{\mathrm{N}_i\mathrm{S}_i} \right).
\end{eqnarray}
\end{subequations}
Combining Eqs. \eqref{eq:muN} with \eqref{eq:IsS}, we can eliminate the spin currents and  imbalances at the NS interfaces to obtain
\begin{eqnarray}
\label{eq:muN2}
\delta\mu_{\mathrm{N}_i}(0)
&=& \frac {eR_\mathrm{N}}2
\left\{
I^s_{\mathrm{F}_i\mathrm{N}_i}
\left[
1-\frac{G_{\rm S}^+}{2(G_{\rm S}^++G_\mathrm{N})}e^{-2L_i/\lambda_{\mathrm{N}}}
\right]
\right.
\nonumber\\
&&
+
\left.
I^s_{\mathrm{F}_{\bar i}\mathrm{N}_{\bar i}}
\frac{G_{\rm S}^+}{2(G_{\rm S}^++G_\mathrm{N})}e^{-(L_1+L_2)/\lambda_{\mathrm{N}}}
\right\},
\end{eqnarray}
where we used the notations $\bar 1=2$ and $\bar 2=1$.

The nonlocal voltage $V_{\mathrm{nl}}$ between F$_2$ and N$_2$ is given by $\mu_{{\rm N}_2}(+\infty)-\mu_{{\rm F}_2}(-\infty)$. Using Eqs.~\eqref{eq:muF} and \eqref{eq:muNx}, we find $V_{\mathrm{nl}}=[\mu_{\mathrm{F}_2\mathrm{N}_2}+p_{\mathrm{F}_2}\delta\mu_{\mathrm{F}_2}(0)]/e$ which,
using Eqs.~\eqref{eq:muF0} and \eqref{eq:GT2a}, may be rewritten as
$V_{\mathrm{nl}}=-(p_{\mathrm{F}_2}/p_\mathrm{F})R_{\mathrm{NF}_i}I^s_{\mathrm{F}_2\mathrm{N}_2}$ where we defined
$R_{\mathrm{NF}_i}= p_{\mathrm{F}}R_{\mathrm{F}}/2+|p_{T_i}|R_{T_i}$ 
and used $p_{T_i}/|p_{T_i}|=p_{\mathrm{F}_i}/p_\mathrm{F}$.
Finally, combining Eqs.~\eqref{eq:muF0}, \eqref{eq:GT2b}, and \eqref{eq:muN2}, we determine the spin current $I^s_{\mathrm{F}_2{\rm N}_2}$ as a function of the injection current $I$ to obtain the nonlocal spin resistance,
\begin{equation}
\label{eq:Rnl}
R_{\mathrm{nl}} = \pm\frac{R_{\mathrm{NF}_1}R_{\mathrm{NF}_2} R_0 e^{-(L_1 + L_2)/\lambda_\mathrm{N}}}{R_{\mathrm{NS}_1} R_{\mathrm{NS}_2} - R_0^2 e^{-2(L_1+L_2)/\lambda_\mathrm{N}}},
\end{equation}
where $R_0 = G_{\rm S}^+/[4G_\mathrm{N}(G_{\rm S}^++G_\mathrm{N})]$ and
$R_{\mathrm{NS}_i} = R_\mathrm{N}/2+R_\mathrm{F}/2+R_{T_i}-R_0 e^{-2L_i/\lambda_N}$.
The overall sign of $R_{\mathrm{nl}}$ depends on whether the ferromagnets are aligned parallel ($+$) or antiparallel ($-$). 

Equation \eqref{eq:Rnl} is the central result of this paper: it predicts a finite nonlocal resistance with no
charge current injected into the superconductor. The exponential dependence $\propto e^{-(L_1+L_2)/\lambda_\mathrm{N}}$ of $R_\mathrm{nl}$ on the lengths of the normal wires clearly shows that it is 
due to the spin transport through the structure. In the case of fully polarized ferromagnets, where 
$|p_{T_i}|=p_\mathrm{F} =  1$ and $R_\mathrm{F}/2+R_{T_i}\rightarrow\infty$, the result \eqref{eq:Rnl} simplifies to $R_\mathrm{nl}= \pm R_0e^{-(L_1 + L_2)/\lambda_\mathrm{N}}$.

A nonloncal spin signal is possible also in the absence of superconductivity. Note that, at $G_{\rm S}^+\gg G_\mathrm{N}$, when the contribution of the superconducting element to the nonlocal signal is negligible,  Eq.~\eqref{eq:Rnl} is similar to the result of Ref. \cite{Taka} up to factors due to a different geometry of the normal part of the FNF spin valve.
In our geometry, a conventional spin valve is realized at temperatures above the superconducting transition temperature $T_c$. We expect the decay lengths of the nonlocal signal within the superconductor to be quite different below and above the transition. At $T\ll T_c$, our results yields $R_\mathrm{nl}\propto e^{-d/\xi_{\rm S}}$ since $G_\mathrm{CAR/EC}$ decay exponentially on the scale ${\xi_\mathrm{S}}$. In contrast, above $T_c$, the nonlocal resistance should be proportional to $e^{-d/\lambda_\mathrm{S}}$, where $\lambda_\mathrm{S}$ is the spin diffusion length of the superconductor in the normal state. Typically, $\lambda_\mathrm{S}$ for Al is $\sim 500 - 1000$ nm and $\xi_{\rm S}$ is $\sim 100 - 300$ nm. Thus, one would expect an abrupt change in the nonlocal resistance when the superconductor transitions to its zero-resistance state.  This effect is possible to demonstrate in experiments similar to Ref.~\cite{Beckmann}. 

The magnitude of the nonlocal resistance that is induced by the injection of pure spin current may be estimated from the nonlocal signal measured in the case of charge current injection. In that case, based on the formalism suggested by Falci \textit{et al}. \cite{Falci}, the nonlocal resistance can be written as
$R^{\mathrm{NSN}}_\mathrm{nl} = ({G_\mathrm{EC}-G_\mathrm{CAR}})/{G_\mathrm{AR}^2}$,
where the conductance due to AR at a single NS interface $G_\mathrm{AR}\gg G_\mathrm{CAR}, G_\mathrm{EC}$ is assumed to be the same for the N$_1$S and N$_2$S interfaces.  
Using the measured $R^{\mathrm{NSN}}_\mathrm{nl}$ as well as estimated values of $G_\mathrm{AR}$ from Cadden-Zimansky \textit{et al.} \cite{cz}
yields a rough estimate of $G_\mathrm{EC}-G_\mathrm{CAR}\sim 0.5 \; \Omega^{-1}$. A lower bound for the nonlocal spin resistance can be obtained if we assume that $G_\mathrm{EC}\gg G_\mathrm{CAR}$. For copper wires with a spin diffusion length $\lambda_\mathrm{N} \sim 1$ $\mu$m and cross section $A_\mathrm{N}$ =100$\times$50 nm$^2$, corresponding to $G_\mathrm{N} \approx 0.3 \; \Omega^{-1}$, the factor $R_0$ in Eq.~\eqref{eq:Rnl} would be of the order of 0.5 $\Omega$ in this case. If $G_\mathrm{CAR}$ and $G_\mathrm{EC}$ are of the same order of magnitude, the nonlocal resistance would likely be much larger.

In summary, we have investigated the nonlocal signal that arises from the  injection of pure spin current into a superconductor.  We have shown that  a finite electrical resistance may be generated entirely due to the nonlocal 
correlations mediated by a superconductor.  In contrast to other recent work, this nonlocal signal arises from spin transport at energies far below the superconducting gap.  Measurements of the nonlocal resistance resulting from  charge  and spin injection on the same sample would allow independent determination of the contributions due to CAR and EC.

Work at Northwestern University was supported by the US NSF under grant No. DMR-1006445. Work at Grenoble was supported through ANR grant ANR-12-BS04-0016-03.

\end{document}